\documentclass[reprint, superscriptaddress, secnumarabic, amssymb, nobibnotes, aps, prl]{revtex4-2}

\usepackage{times,xspace}
\usepackage{graphicx,graphics,color,epsfig}
\usepackage{amsmath}
\usepackage{amssymb}
\usepackage{amsfonts}
\usepackage{amsfonts}
\usepackage{epstopdf}
\usepackage[T1]{fontenc}
\usepackage{amsmath}
\usepackage{amssymb}
\usepackage{siunitx}
\usepackage{braket}
\usepackage{xcolor}
\usepackage{bbm}
\usepackage{color}
\usepackage{float}
\usepackage[colorlinks=true]{hyperref}

\hypersetup{
    bookmarks=true,         
    unicode=false,          
    pdftoolbar=true,        
    pdfmenubar=true,        
    pdffitwindow=false,     
    pdfstartview={FitH},    
    pdftitle={Superconducting properties of Re$_{2}$X (X = Hf, Zr)},    
    pdfauthor={M. Mandal, R. P. Singh},      
    pdfsubject={},   
    pdfcreator={},   
    pdfproducer={}, 
    pdfkeywords={} {} {}, 
    pdfnewwindow=true,      
    colorlinks=true,       
    linkcolor=blue, 
    citecolor=blue,        
    filecolor=magenta,      
    urlcolor=blue           
} 

\usepackage[normalem]{ulem}

\newcommand{\figref}[1]{Fig.~\ref{#1}}

\renewcommand{\approx}{\simeq}

\def\be{\begin{equation}}
\def\ee{\end{equation}}
\def\bea{\begin{eqnarray}}
\def\eea{\end{eqnarray}}

\begin{document}

\title{Time-reversal symmetry breaking in frustrated superconductor Re$_2$Hf}
\author{Manasi~Mandal}
\affiliation{Department of Physics, Indian Institute of Science Education and Research Bhopal, Bhopal, 462066, India}
\author{Anshu~Kataria}
\affiliation{Department of Physics, Indian Institute of Science Education and Research Bhopal, Bhopal, 462066, India}
\author{Chandan~Patra}
\affiliation{Department of Physics, Indian Institute of Science Education and Research Bhopal, Bhopal, 462066, India}
\author{D.~Singh}
\affiliation{ISIS Facility, STFC Rutherford Appleton Laboratory, Harwell Science and Innovation Campus, Oxfordshire, OX11 0QX, UK}
\author{P.~K.~Biswas}
\affiliation{ISIS Facility, STFC Rutherford Appleton Laboratory, Harwell Science and Innovation Campus, Oxfordshire, OX11 0QX, UK}
\author{A.~D.~Hillier}
\affiliation{ISIS Facility, STFC Rutherford Appleton Laboratory, Harwell Science and Innovation Campus, Oxfordshire, OX11 0QX, UK}
\author{Tanmoy Das}
\email[]{tnmydas@iisc.ac.in}
\affiliation{Department of Physics, Indian Institute of Science, Bangalore-560012, India.}
\author{R.~P.~Singh}
\email[]{rpsingh@iiserb.ac.in}
\affiliation{Department of Physics, Indian Institute of Science Education and Research Bhopal, Bhopal, 462066, India}

\date{\today}
\vspace{0.3cm}

\begin{abstract}
Geometrical frustration leads to novel quantum phenomena such as the spin-liquid phase in triangular and Kagom\'e lattices. Intra-band and inter-band Fermi surface (FS) nesting can drive unique superconducting (SC) ground states with $d$-wave and $s^{\pm}$ pairing symmetries, respectively, according to the criterion that the SC gap changes sign across the nesting wavevector. For an odd number of FSs, when multiple inter-band nesting is of comparable strength, the sign-reversal criterion between different FS sheets can leads to frustration, which promotes novel SC order parameters. Here we report the experimental observation of a time-reversal symmetry breaking pairing state in Re$_2$Hf resulting from FS nesting frustration. Furthermore, our electronic specific heat and transverse-field $\mu$SR experiments suggest a fully gaped pairing symmetry. The first-principle electronic structure calculation reveals multiple Fermi surface sheets with comparable inter-band nesting strength. Implementing the {\it ab-initio} band structure, we compute spin-fluctuation mediated SC pairing symmetry which reveals a $s+is'$-pairing state - consistent with experimental observations. Our investigation demonstrates a novel SC state which provides a putative setting for both applied and fundamental study. \\
\end{abstract}
\date{\today} 

\maketitle



Conventional superconductivity is mediated by electron-phonon coupling, which induces an attractive interaction that results in a unique ground state of isotropic and fixed-sign pairing symmetry. The attractive interaction leaves little room for exotic pairing symmetries. On the other hand, while unconventional superconductivity is yet to be fully understood, the possibility of having novel and exotic pairing symmetries due to the interplay of structural symmetries and Fermi surface (FS) topology make it an attractive topic of research \cite{SigristRMP}.

The theory of spin-fluctuation mediated unconventional superconductivity requires a pairing symmetry that must change sign across the Fermi surfaces (FSs) related to the nesting wave vector \cite{SCcuprates,SCcuprates2,SCcuprates3,SCcuprates4,SCcuprates5,SCcuprates7,SCcuprates8,SCcuprates9,SCcuprates10,SCHF,Othermaterials,Othermaterials2}. In a single FS sheet, if the pairing symmetry changes sign between different parts of the same FS, it must go through a zero gap state. This gives rise to nodal superconductivity; for example: $d$-wave gap symmetry in cuprates \cite{SCcuprates,SCcuprates2,SCcuprates3,SCcuprates4,SCcuprates5,SCcuprates7,SCcuprates8,SCcuprates9,SCcuprates10}, in some of the heavy-fermions \cite{SCHF,SCHF2,SCHF3,SCHF3,SCHF4,SCHF5}, and other compounds \cite{Othermaterials,Othermaterials2,ThCoCl}. In a multiband SC, if the FS nesting occurs between different FS sheets, the SC gap may possess an opposite sign between different FS sheets but a single sign on a single FS sheet. Such solutions lead to the $s^{\pm}$ pairing state, which is fully gaped (nodeless), as seen in pnictides \cite{SCpnictides,SCpnictides2,SCpnictides3,SCpnictides4} and proposed in some other systems \cite{SCHF3,Otherspm}.

\begin{figure}[ht!]
\centering
\includegraphics[width=0.45\textwidth]{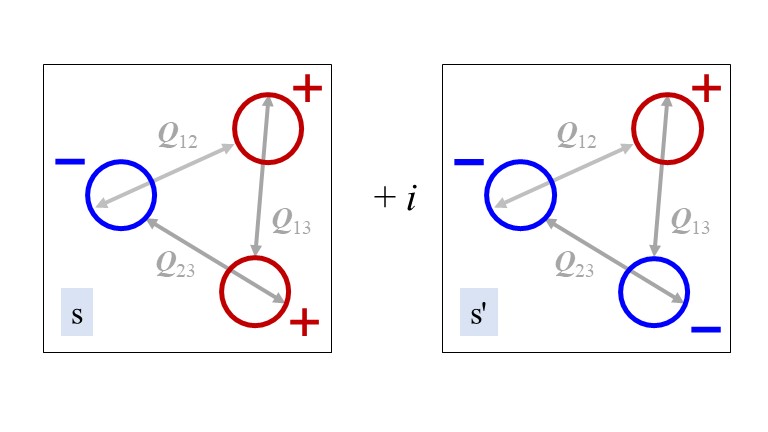}
\caption{\label{Fig1} A schematic demonstration of frustrated pairing symmetry in a three FSs (circles). Blue and red color denote different signs of the SC gap. $Q_{ij}$ are the assumed inter-band nesting vectors between the $i$ and $j^{\rm th}$ FS sheets, with comparable nesting strength. Each nesting promotes a sign reversal of the SC gap between nested FSs, hence two degenerate $s$ and $s'$ pairing symmetries arise. The ground state is the linear superposition of the two solutions, which is a $s+is$ pairing symmetry. 
}
\label{theory_fig1}
\end{figure}

It is interesting to ask what happens when there exists an odd number of FS sheets (say three), and the pairing interaction promotes an inter-band sign-reversal pairing symmetry. How do we fix the sign of the gap on the third FS sheet, as schematically shown in \figref{theory_fig1}. For example, when the three inter-band FS nestings are of comparable strengths and promote an inter-band $s^{\pm}$ pairing state, any two FSs possess opposite SC gaps, while the third FS has the equal probability of having $\pm$ SC gap. In such a frustrated SC phase, the lowest ground state can be a superposition state of both gaps, which gives a time-reversal breaking $s\pm i s$ pairing state. So far, time-reversal symmetry breaking (TRSB) superconductivity have been reported in a very few materials, e.g. Sr$_{2}$RuO$_{4}$ \cite{SrRuO,SrRuO_2} and (U,Th)Be$_{13}$ \cite{UBe_1,UBe_2,UBe_3,UBe_4}, (Pr,La)(Ru,Os)$_{4}$Sb$_{12}$ \cite{PrSb_1,PrSb_2}, PrPt$_{4}$Ge$_{12}$ \cite{Ge}, LaNiGa$_{2}$ \cite{Ga}, Lu$_{5}$Rh$_{6}$Sn$_{18}$ \cite{LuRhSn}, Re$_{6}$X \cite{ReZr,ReHf,Ti,Ti2}, LaNiC$_{2}$ \cite{LaNiC2}, La$_{7}$(Ir/Rh)$_{3}$ \cite{La7Rh, La7Ir} and very recently in Sr$_{x}$Bi$_{2}$Se$_{3}$ \cite{SrBiSe}, 4Hb-TaS$_{2}$ \cite{TaS2} and Ba$_{1-x}$K$_{x}$Fe$_{2}$As$_{2}$ \cite{FeAs}. In these materials, the TRSB superconductivity is often stabilized by mainly ferromagnetic fluctuations or spin-orbit coupling (SOC). 

In this work, we report the evidence of TRSB SC ground state resulting from a novel mechanism of FS nesting frustrations in Re$_{2}$Hf. This constitutes the first direct experimental observation of multiband superconductivity exhibiting frustration.

\begin{figure}[ht!]
\includegraphics[width=1.0\columnwidth]{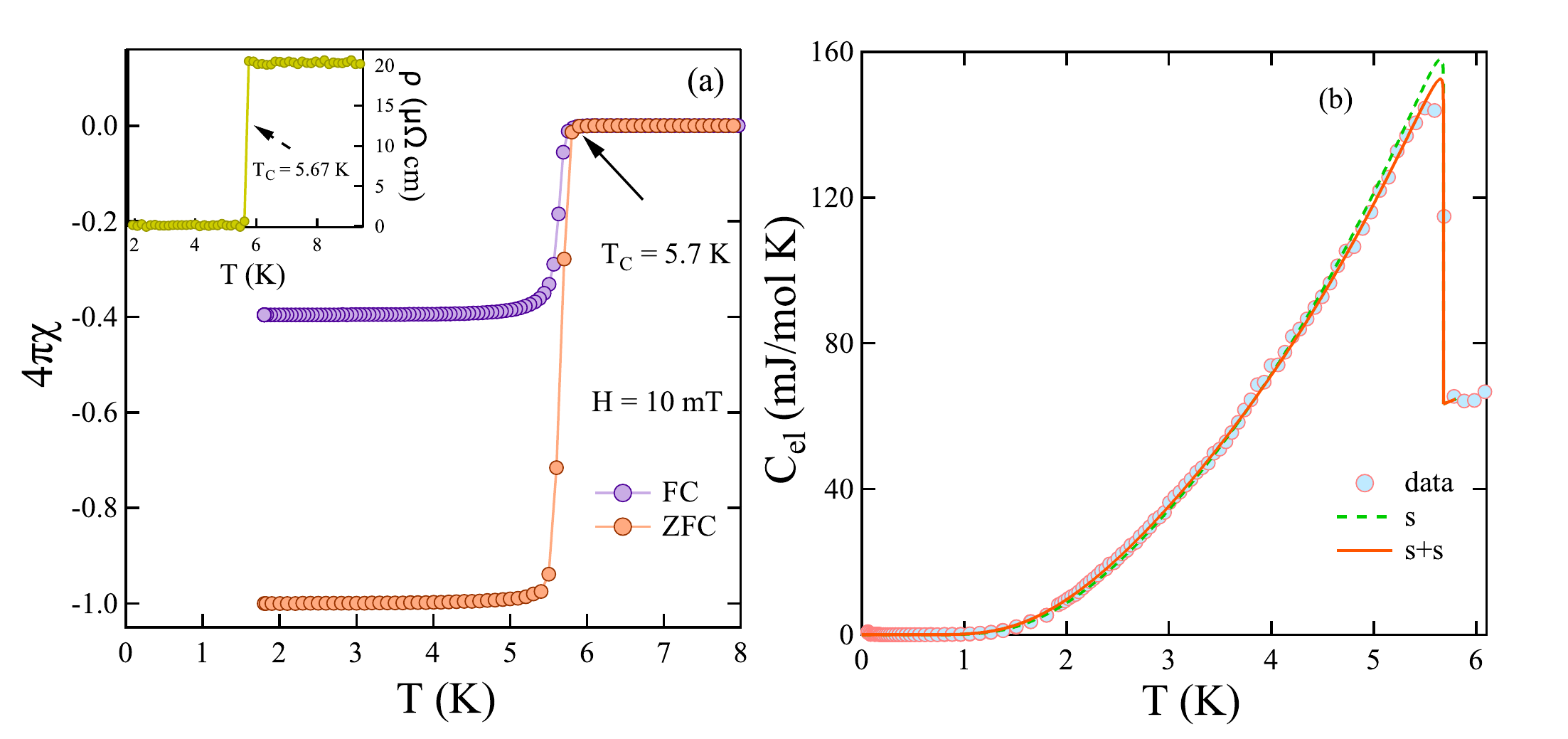}
\caption {\label{Fig2} (a) The temperature dependence of the magnetic susceptibility at 10 mT applied field and inset highlights the zero resistivity. (b) The temperature dependence of the electronic specific heat, C$_{el}$ in the superconducting state fitted with $\alpha$ model. Inset highlights the difference between the observed data and the theoretical model.}
\end{figure}


Re-based binary compounds (Re-X) form in centrosymmetric and non-centrosymmetric crystal structures depending on the Re/X stoichiometry ratio and most of these binary compounds are superconductors. Re$_{2}$Hf is another member of this family, forming a C14 Laves phase with a centrosymmetric crystal structure. Hexagonal C14-type Laves phase is a form of well-known Laves phases with the general composition of AB$_{2}$, and the B atoms often dominate the electronic properties of the Laves phases \cite{laves}.

The Re$_{2}$Hf sample was prepared via the standard arc melting method. We characterized the compound by magnetization, resistivity, and specific heat measurements. These measurements confirm that Re$_{2}$Hf is a type-II superconductor with T$_{C}^{mag}$ = 5.7(1) K (\figref{Fig2}(a)) and lower critical field, H$_{C1}$(0) field = 12.7(1) mT (SI). The temperature variation of the upper critical field, H$_{C2}$(T), shows concave nature in the high field region, which is consistent in resistivity, magnetization, and specific heat measurements (as shown in SI). Ginzburg-Landau (GL) model give a reasonable fitting only to the low field region. In contrast, a two-gap model is required to obtain a good fitting of the data in the whole region (dash line gives a two-gap fit), indicating the possible multiband superconductivity in Re$_{2}$Hf compound.

\begin{figure}[ht!]
\includegraphics[width=1.0\columnwidth]{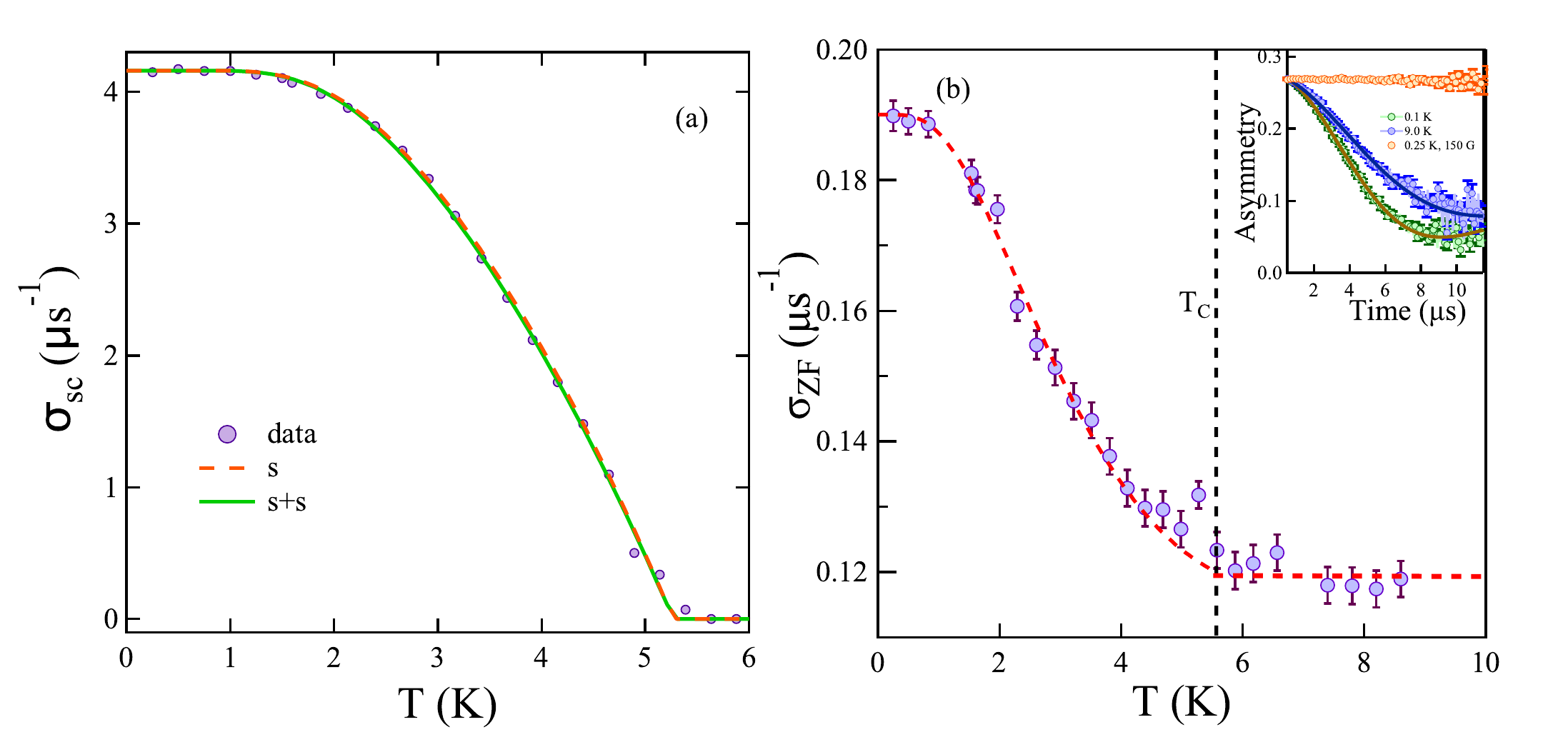}
\caption {\label{Fig3} (a) The temperature dependence of $\sigma_{sc}$  is fitted with different models. (b) Temperature dependence of $\sigma_{\mathrm{ZF}}$ and inset shows the time evolution of the spin polarization of muons under zero-field conditions.}
\end{figure}

The exact SC gap nature is investigated using specific heat measurements. The electronic contribution to the specific heat was fitted using different models \cite{alpha}, as shown in \figref{Fig2}(b). It is visible that the $s+is'$ model fits better over the whole temperature range than the $s$ and $s+id$ models. The good agreement of the $s+is$ model with observed data indicates the presence of multiple nodeless gaps on the FS. The field variation of the Sommerfield coefficient, $\gamma$, in the vortex state show a nonlinear behavior as shown in {\it SI}. $\gamma$ varies linearly with the external field, $H$ up to a certain crossover field of 0.07*H$_{C2}(0)$ whereas it is well fitted with H$^{0.5}$ over the whole region. The black dashed line guides the eye to two linear fits with different slopes, which can correspond to the two isotropic gaps on the FSs. A similar behavior of the Sommerfield coefficient was observed for popular multigap superconductor such as MgB$_{2}$ \cite{MgB2_1,MgB2_2}, 2H-NbSe$_{2}$ \cite{NbSe2_1}. Such a nontrivial nature of $\gamma$  suggests the presence of an unconventional single- particle energy gap, as predicted by the theoretical results \cite{sas,line_nodes,line_nodes2}.\\
The gap structure of the SC state is further probed by the transverse-field $\mu$SR  measurements. The normal state spectra show a homogeneous field distribution throughout the sample, with a weak depolarization arises from the dipolar nuclear field. In contrast, the SC state spectra show a strong depolarization, indicating the formation of inhomogeneous field distribution in the flux line lattice (FLL) state {\it SI}.\\
The temperature dependence of the depolarization arising due to the field variation across the flux line lattice ($\sigma_{\mathrm{sc}}$) is displayed in \figref{Fig3}(a). The temperature dependence of $\sigma_{sc}$ is seen nearly constant below $\approx$ $T_{C}$/3, indicating the absence of nodes in the SC energy gap at the FS. We have fitted the data with different models, while only a two gaps scenario with both gaps being nodeless gives good fitting over most of the region. The details of fitting functions are summarized in {\it SI}. This suggests the dominating $s$-wave behavior in the compounds with an isotropic or nearly isotropic gap in the electronic density of states at the Fermi level. Two gap model yields a value of the BCS ratio $\frac{\Delta(0)}{k_{B}T_{C}}$ as 1.89 and 1.82 with a weighting factor 0.21 to the second gap. While both single and two-gap $s$-wave models yield good agreement but the two-gap fitting covers the whole temperature region with better goodness of fit. The gap values are slightly higher than those estimated from specific heat values due to the applied field during measurement. The calculated value of  $\frac{\Delta(0)}{k_{B}T_{C}}$ is higher than the BCS predicted value of 1.76, indicating deviations from the weak coupling BCS pairing scenario. The fitting with local London approximation in the clean limit yields $\lambda(0)$ = 1397(10)\AA.\\
To further investigate the occurrence of TRS breaking in Re$_{2}$Hf, a systematic zero-field $\mu$SR was performed at different temperatures. The inset of \figref{Fig3}(b) shows the ZF-$\mu$SR spin relaxation spectra, where there is a clear change in the relaxation behavior of the spectra recorded above (T = 9 K) and below (T = 0.1 K) T$_{C}$. This indicates the presence of spontaneous magnetic fields in the SC state. We have fitted ZF-$\mu$SR data by a damped Gaussian Kubo-Toyabe (KT) function \cite{KT1} with a background contribution associated with the muon stopping in the silver sample holder. The temperature dependence of the Gaussian relaxation rate, $\sigma_{\mathrm{ZF}}$ is shown \figref{Fig3}(b). $\sigma_{\mathrm{ZF}}$ gradually increases below the superconducting transition temperature (dotted line guides to the eye), indicating a possible TRS breaking signal. 
We have also performed LF-$\mu$SR measurements at 0.25 K of 15 mT, shown by orange flat asymmetry spectra in \figref{Fig3}(b). The applied field decouples the static internal field, excluding the possibility of an impurity-induced relaxation. The calculated value of the spontaneous field, B$_{int}$ in the SC region is 0.116 mT. The appearance of such spontaneous fields in the SC state provides strong evidence for a TRS broken pairing state. The internal field value is highest among all the TRS breaking superconductors, suggesting a distinctive and novel SC state in this compound.

{\it Electronic properties}. From the general physics of Leaves phase elements as Re$_2$Hf is, it is expected that the electronic structure of Re plays a crucial role in superconductivity\cite{laves,Re}. Re$_{2}$Hf belongs to the P6$_{3}$/$mmc$ space group (No. 194) and has the hexagonal point group of $D_{6h}$. We compute the electronic structure using the Vienna Ab-initio Simulation Package (VASP) \cite{VASP}, and the Perdew-Burke-Ernzerhof (PBE) form for the exchange-correlation functional \cite{PBE}. The lattice parameters are allowed to relax, and the values are found to be a = 5.29 ${\rm A}^{o}$, and c/a = 1.63, close to the experimental value \cite{book}. To deal with the strong correlation effect of the $d$-electrons of Re atoms, we employed the LDA+$U$ method with the standard database value of $U = 2.4$ eV. The spin-polarized calculation is performed to confirm that the material is nonmagnetic. Finally, the calculations are repeated with and without the spin-orbit coupling. 

\begin{figure}[t]
\centering
\includegraphics[width=0.45\textwidth]{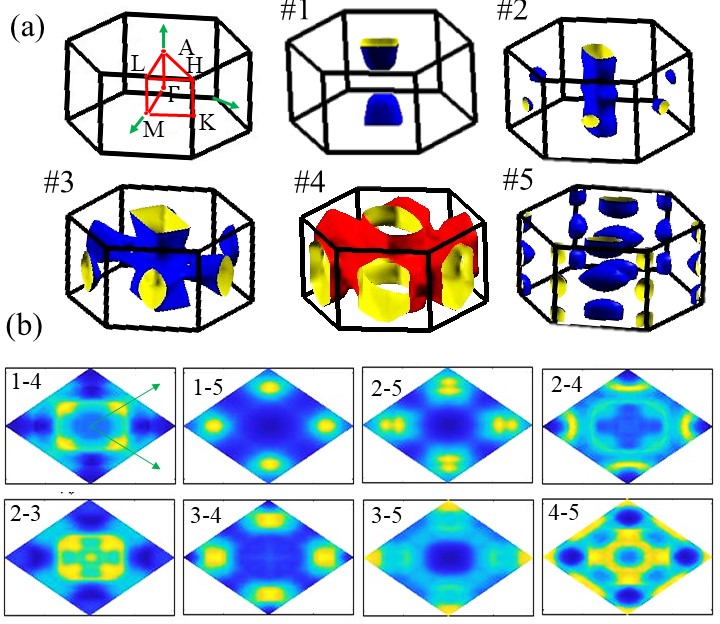}
\caption{{\bf Theoretical results of the FS topology and FS nestings.}(a) Computed FS topologies for different bands in the first Brillouin zones for calculations with SOC. The blue and red color on the FSs denote the sign of the computed pairing symmetry, while yellow color is for visualization. (b) Computed static bare susceptibilities are shown for several representative inter-band components, which are dominant in this material. Blue to yellow color denote minimum to maximum value of the susceptibility.}
\label{theory_fig}
\end{figure}

There are five FS sheets, as shown in \figref{theory_fig}(a) for the SOC split bands. All FSs show substantial three-dimensional behavior. Among all the FSs, FS 4 has the largest area and then FS 3 and so on. For all bands, we find intra-band FS nesting is much weaker than the inter-band components, and expectedly, the FS nesting strength between FS 4 to others is stronger while those among the other FSs are equivalent, causing frustrations in the nesting profile, see \figref{theory_fig}(b). Without SOC, the topology of the FSs remains similar, but two additional small FSs appear. The following calculations of the SC pairing symmetry are repeated for the SOC-free band structures, and the results of the $s+is'$ pairing symmetry are found to be robust.

{\it Spin-fluctuation mediated superconductivity.} We consider a multi-band Hubbard model with intra- and inter-band onsite interactions $U$, and $V$, respectively. The density-density fluctuation mediated pairing interaction $\Gamma({\bf q})$ is obtained within the random-phase-approximation (RPA) by summing the so-called bubble and ladder diagrams\cite{SCcuprates,SCpnictides,SCHF,SCrepulsive}, and the result is:
\begin{eqnarray}
\Gamma({\bf q})&=&\frac{1}{2}\big[3U_{s}{\chi}_{s}({\bf q}){U}_{s} - {U}_{c}{\chi}_{c}({\bf q}){U}_{c} + {U}_{s}+{U}_{c}\big].
\label{singlet}
\end{eqnarray}
The subscripts `s' and `c' denote spin and charge fluctuation channels, respectively. ${U}_{s/c}$ are the onsite interaction tensors for a spin and charge fluctuations, respectively, whose diagonal terms involve intra-band Hubbard $U$ and the off-diagonal components give the inter-band Hubbard $V$. ${\chi}_{s/c}$ are the RPA spin and charge susceptibilities computed by directly including the DFT band structures. The details of the formalism are given in {\it SI}.

We compute the eigenvalue and eigen functions of the pairing interaction $\Gamma({\bf q}={\bf k}-{\bf k}')$ on the 3D FS by solving the following equation:
\begin{eqnarray}
{\bf \Delta}({\bf k})= -\lambda\sum_{{\bf k}'}{\bf \Gamma}({\bf k}-{\bf k'}){\bf \Delta}({\bf k'}).
\label{SC2}
\end{eqnarray}
${\bf \Delta}({\bf k})$ is a $N$-component eigenvector at each ${\bf k}$, denoted by ${\bf \Delta}({\bf k})=\{\Delta_i({\bf k})\}$, where $i=1-N$ with $N$ being the total number of bands. 
To consider the intra- and inter-band nestings on equal footings, we construct a pairing potential matrix ${\bf \Gamma}$ of dimension $N'\times N'$, where $N'= N\times N_{k}$ with $N_{k}$ being the total number of coarse-grained Brillouin zone k-points. $\lambda$ is the overall pairing eigenvalue (proportional to the SC coupling strength). Since the pairing potential is repulsive here, the highest {\it positive} eigenvalue $\lambda$, and the corresponding pairing symmetry can be shown to govern the lowest Free energy value in the SC state.\cite{SCrepulsive} 

We numerically solve Eq.~\eqref{SC2}, and find several (nearly)  degenerate solutions. We present one of the solutions by projecting the eigen state ${\bf \Delta}({\bf k})$ onto different band basis $\delta_{i}({\bf k})$ in \figref{theory_fig}(a) by blue to red color (positive to negative value of the SC gap) on the FSs. We make several important observations. (i) All the degenerate gap solutions yield a negative gap on FS \#4, while the sign of the gap is frustrated among the remaining four FSs. This is consistent with the fact that the FS \#4 has the largest volume and has comparable inter-band nestings with all other FSs. Therefore, the SC solution is favoured for a sign reversal in all other bands with respect to FS \#4. (ii) The other nesting channels have variable strength and nesting wavevectors. Several dominant (static) spin-fluctuation profiles are shown in \figref{theory_fig}(b). We observe that there are several inter-band nesting profile which are nearly degenerate. For example, the nestings between FSs \#2-\#4 are very much degenerate with that between \#3-\#4, \#1-\#5, and \#2-\#5. Therefore, for the negative gap in FS \#4, the \#1-5 and \#2-\#5 nestings frustrate the sign on the FS \#5 with respect to \#1, \#2. Similarly, the similar nesting profiles between bands \#1-\#4, \#2-\#3, and \#4-\#5 tend to frustrate the SC gap's sign between FS \#1 and \#5. (iii) In all cases, we find that the intra-band FS nesting strength is considerably small. This plays a key role in the absence of any ${\bf k}$-space anisotropy of the gap within a given band. 

{\it Discussion of the SC gap symmetry.} The origin of the TRS breaking complex pairing eigen state is understood as follows. We express the gap functions as $\Delta_i({\bf k})=\eta_ig_i({\bf k})$, where $\eta_i$ is the gap value and $g_{i}({\bf k})$ incorporates the ${\bf k}$-space anisotropy, and $i$ is the band index. The sign-reversal of the gaps can now occur either in $\eta_i$ between different bands ${\rm sgn}[\eta_i]\ne {\rm sgn}[\eta_{j\ne i}]$, and/or ${\rm sgn}[g_i({\bf k})]\ne {\rm sgn}[g_j({\bf k}+{\bf q})]$ for $i=j$ (intra-band) or $i\ne j$ (inter-band), where ${\bf q}$ is the nesting a vector. The hexagonal group $D_{6h}\otimes T$, where $T$ is the time-reversal symmetry, is split into the $A_{1g}$ and $E_{2g}$ irreducible representations.\cite{Agterberg} This implies that $g_i({\bf k})$ can have $s$-wave or $d$-wave symmetries. 

(i) In the limit when the inter-band interactions are much smaller than the intra-band components, Eq.~\eqref{SC2} decouples to $N$-separate eigenvalue equations in the momentum space. Given that $\lambda>0$, and $\Gamma>0$, the only allowed solutions of Eq.~\eqref{SC2} are those for which each $g_i({\bf k})$ component changes sign across the corresponding nesting vector. This means $g({\bf k})\in E_{2g}$. There is little possibility in this case to obtain any other exotic order parameter such as a TRSB complex order parameter. (ii) In the other limit of $\Gamma_{i\ne j,j}>>\Gamma_{ii}$, generally both $A_{1g}$ and $E_{2g}$ symmetries may arise: $A_{1g}$ is preferable when the sign reversal occurs in $\eta_i$ between two bands due to inter-band nesting without a preferential wavevector or if there are multiple nestings of comparable strengths causing frustration (our case). This gives a $s^{\pm}$-pairing symmetry. $E_{2g}$ symmetry can arise here when the inter-band nesting promotes a sign reversal in both $\eta_i$ and $g_i({\bf k})$. In both cases, one may obtain a TRSB combination such as a $s+is$, or $d+id$ for the reasons discussed below. (iii) In the intermediate region where both intra-band and inter-band nestings are comparable, both $A_{1g}$ and $E_{2g}$ symmetries are allowed, and since these two irreducible representations do not mix, a stable solution would require a TRSB combination, i.e. an $s+id$ order parameter.   
Our numerical simulation predicts an $s+is'$ pairing symmetry which belong to point (ii) above. Here the origin of an $s+is'$ combination interesting. Because of the $s$-wave nature in each band, $g_i({\bf k})=\pm 1$. There arises an internal rotational symmetry $SO(N)$ of the components of the order parameter ${\bf \Delta}=\{\eta_i\}$ as, for example, $\eta_1\rightarrow \eta_2$, $\eta_2\rightarrow \eta_3$, ..., $\eta_N\rightarrow \eta_{1}$. Therefore, we can Fourier transform the order parameters as
\begin{equation}
\bar{\eta}_{p}=\frac{1}{\sqrt{N}}\sum_{n=0}^{N-1}\eta_n\exp{\left(i\frac{2\pi p}{N}n\right)},
\label{Eq:FTgap}
\end{equation}
where $n$ runs over the number of bands and $p\in \mathbb{Z}$. Clearly, although the pairing amplitude $\eta_i$ for each band is real, the Fourier component $\bar{\eta}_{p}$ is complex for $p\ne 0$, and breaks the TRS. The phases of the order parameter is clearly dictated by the total number of bands. For our five band model, we have the order parameters which takes the form $\bar{\eta}_1= 1/\sqrt{5}\sum_{n=0}^4\eta_n\exp(i2\pi n/5)$. This multiband order parameter is consistent with the multi-gap behavior we observed in our experiments.


To summarise, we have investigated Re$_{2}$Hf. Zero and longitudinal-field $\mu$SR data reveal the presence of spontaneous static magnetic fields below T$_{C}$, confirming that time-reversal symmetry is broken in the superconducting state. Detail theoretical work suggests the FS nesting frustration in the superconducting ground state leads to time-reversal symmetry breaking. This is the first experimental observation of frustrated multiband superconductivity. This work paves the way for further studies of many superconductors in the Re$_{2}$X family in the hunt for unconventional behavior. Apart from the local magnetic moment corresponding to the TRSB SC as observed here, such an exotic order parameter has other interesting experimental signatures \cite{spispnictide,TRSBSC_General,TRSBSC_General2,TRSBSC_General3,spis_defects}, and also induces novel collective \cite{TRSBSC_mode,TRSBSC_mode2} and topological excitations \cite{TRSBSC_topological,TRSBSC_topological2,TRSBSC_topological3}.\\

 R.~P.~S.\ acknowledges the Science and Engineering Research Board, Government of India for the Core Research Grant CRG/2019/001028 and Financial support from DST-FIST Project No. SR/FST/PSI-195/2014(C) is also thankfully acknowledged. We thank ISIS, STFC, United Kingdom, for the muon beam time.TD's work is supported by the NSM project, and facilitated by the S.E.R.C. supercomputing facility at I.I.Sc.

\end{document}